\documentclass{jaa}

\usepackage{graphicx}
\usepackage{hyperref}

\AtBeginDocument{}

\begin{document}\sloppy

\title{Effect of intra-channel baseline migration on  \\ 
the measured visibility and spatial power spectrum \\ }


\author{Magendran S\textsuperscript{1}}
\affilOne{\textsuperscript{1} Raman Research Institute, Bengaluru 560080, India.\\}


\twocolumn[{

\maketitle

\corres{magendran@rri.res.in}


\begin{abstract}
The channel-to-channel migration of radio interferometric baselines for the same antenna separation causes a flat spectrum source that should have remained in the zeroth delay (line-of-sight) mode to become centered around a higher mode - the geometric delay for that particular antenna separation with a spread (spill-over) of the order of reciprocal bandwidth. While in principle an errorless gridding interpolation can remove inter-channel migration, intra-channel baseline migration exists due to the non-zero width (resolution) of the instrument's spectral channel arising from the finite-period integration in a DFT cycle. Here for the first time, we analyze this effect and quantify it using the case of a flat-spectrum point source. We find that the visibility undergoes an attenuation, the extent of which depends on the auto-correlation of the window function used for DFT and is more towards the lower-end of the band. This causes the spill-over of the delay mode to extend beyond the reciprocal bandwidth order. \\ 
\end{abstract}

\\ \\  \\ \\
}]



\volnum{000}
\year{2018}
\pgrange{1--}
\setcounter{page}{1}
\lp{11}

\section{Introduction}
The mode-mixing effects due to inter-channel migration have been studied extensively in literature in the context of redshifted 21-cm $H_{1}$ signal. See Vedantham H. {\em et al.} (2012); Morales M. F. {\em et al.} (2012); Thyagarajan N. {\em et al.} (2013); Thyagarajan N. {\em et al.} (2015a, 2015b) and references therein. Inaccuracies in the gridding interpolation due to finiteness of the visibility coverage (both in the number of samples and the sample spacing) cause post-gridding residual jitter in the visibility for a fixed baseline as a function of frequency. As we go to higher antenna separation $(x)$, the channel-to-channel migration ($\Delta u = x \Delta f / C$) increases and the jitter  fluctuates more rapidly from one channel to the next (section 4.1 in Vedantham H. {\em et al.} 2012). This results in mode leakage over a wedge-like region in the instrument k-space where the extent of leakage into higher line-of-sight wave numbers (delay) is proportional to the transverse wave number (baseline). We'll use the terms mode and wavenumber interchangeably.  \\

  While this jitter effect caused by inter-channel migration needs to be addressed by optimising the gridding operation to isolate the highly fluctuating modes (of the gridded visibility with respect to baseline and frequency) not occupied by the true visibility, in this paper we focus on intra-channel baseline migration whose mitigation requires making the channel width approach zero, possibly with a polyphase filter bank with large number of taps in the FIR filters (as we'll see in section \ref{subsec:PFB}). To the best of the author's knowledge this effect has not been treated so far in the literature. It is even considered that there is no need/incentive to improve the spectral channel resolution beyond the need to localise RF interference (see for example section 3.2 in Marthi V. R. (2017)). \\
  
  In section \ref{sec:Section2}, we introduce the micro-tone spectral representation that is required to understand and interpret the intra-channel baseline migration and its consequences. We show how the finite integration time in DFT affects the FX-correlated,  per-channel visibility from a source pixel. We then quantify this artifact for the flat-spectrum source. In section \ref{sec:Section3}, we consider how the spatial power spectrum is affected. \\

\section{Intra-channel migration} \label{sec:Section2}

\subsection{Micro-tone representation  \label{subsec:utone}}

The analog power spectral density (intensity as a continuous function of frequency) from a source pixel can be represented as accurately as needed by specifying it at discrete micro-tones, whose spacing, $\epsilon$ can be made as small as needed. The pixel's double-sided power spectral density and auto-correlation, real and even functions in frequency and delay respectively are:

\begin{eqnarray}
P_{src}(f) &=& \sum_{m=-\infty}^{\infty} P_{m}.\{ \delta(f-\epsilon m) \} \label{eq:p_analog}  \\  \nonumber \\
R_{src}(\tau) &=& P_{0} + \sum_{m=1}^{\infty} 2P_{m}.\cos(2\pi \epsilon m\tau) \label{eq:r_analog} 
\end{eqnarray}
\\

\noindent where $P_{m}$ is the source's power spectral density per unit microtone spacing at the microtone frequency $m\epsilon$. Let the system bandwidth, $f_{sys}$ be in the interval $(f_{start}, \: f_{end}]$ and $f_{res}$ be the DFT resolution. $f_{start}$ is chosen to be an integral multiple of $f_{res}$ and the first spectral channel is centered at $(f_{start} + f_{res})$.

\begin{eqnarray}
f_{res} &=& \sigma \epsilon \label{eq:f_res} \\
f_{start} &=& K.f_{res}  \label{eq:f_start} \\ \nonumber \\
f_{sys} &=& (f_{end} - f_{start}) \nonumber \\
&=& (B_{max}+0.5).f_{res}  \label{eq:f_sys}
\end{eqnarray}
\\

The interferometer measures the received radiation electric field as a function of aperture position and time,  

\begin{eqnarray}
E(x, t) &=& \sum_{m=-(K+B_{max}+0.5)\sigma}^{(K+B_{max}+0.5)\sigma} \{ E_{m}(x).\exp(\jmath 2\pi \epsilon m t) \} \label{eq:E_x_t}\nonumber \\ 
\end{eqnarray}
\\

\noindent $E_{m}(x)$ is the field spectrum for the microtone frequency $f = m\epsilon$ seen at the aperture position $x$. The field spectral samples at positive and negative frequencies are complex conjugates due to $E(x, t)$ being real (i.e., $E_{-m}(x) = E^{*}_{m}(x)$). Band-pass filtering for $f_{start} < |f| \leq f_{end}$ ensures that $E_{m}(x)$ is non-zero only for $(K+0.5)\sigma < |m| \leq (K+B_{max}+0.5)\sigma$.  \\

$E(x, t)$ is sampled at a period of $T_s = (2f_{end})^{-1}$. To analyse baseline migration  which happens at the original radio frequency, we use direct RF (rather than the usual down-converted intermediate frequency (IF)) sampling. If analysing for a system with IF down-conversion, the signal band needs to be mapped (one-to-one) back to the original RF band while the spectral resolution is what is achieved in the IF band. Let $N_\epsilon$ be the number of samples contained in one microtone period ($T_\epsilon = \epsilon^{-1}$). It is also the number of microtones contained in the band $|f| \leq f_{end}$.  

\begin{eqnarray}
N_{tot} &=& 2 f_{end}/f_{res} = 2 (K+B_{max}+0.5) \label{eq:N_tot} \\
N_\epsilon &=& T_\epsilon / T_s = N_{tot} \sigma + 1 \label{eq:N_e} 
\end{eqnarray}
\\ 

\begin{eqnarray}
E(x, t=n T_s) &=& \sum_{m=(K+0.5)\sigma+1}^{0.5 N_{\epsilon}} \{ E_{m}(x).\exp(\frac{\jmath 2\pi m n}{N_\epsilon}) \nonumber \\
&+& E^{*}_{m}(x).\exp(\frac{-\jmath 2\pi m n}{N_\epsilon}) \} \label{eq:E_x_t_2} \nonumber \\ 
\end{eqnarray}
\\

In FX mode, the detected field is first transformed into spectral channels/bins using DFT and then spatially correlated over the aperture. The observation time for each DFT cycle is $T = {f_{res}}^{-1} = ({\sigma\epsilon})^{-1} = \sigma^{-1}T_{\epsilon} = N_{tot} T_s$. The $b^{th}$ spectral bin corresponds to a bin-center frequency of $f_b = b f_{res}$, for $|b| \leq (K+B_{max})$. The DFT spectrum at the position $x$ is:

\begin{eqnarray}
E(x, b) &=& \frac{T_s}{T} \sum_{n=0}^{N_{tot}-1} \{ h(t=nT_s) \label{eq:E_b} \nonumber \\
&.& E(x, t=nT_s).\exp\left(\frac{-\jmath 2 \pi b n}{N_{tot}}\right) \}  \nonumber \\ \\ \nonumber \\  
&=& \frac{\sigma T_s}{T_{\epsilon}} \sum_{n=0}^{N_{\epsilon}-1} \{ h(t=nT_s) \nonumber \\
&.& E(x, t=nT_s).\exp\left(\frac{-\jmath 2 \pi (b\sigma) n}{N_\epsilon}\right) \}  
\nonumber \\ \label{eq:E_b2}  
\end{eqnarray}
\\

The real-valued window function $h(t)$ to control spectral leakage between channels, is non-zero for $0 \leq t \leq T$. It is zero-padded till $|t|=0.5T_\epsilon = 0.5 \sigma T$ and replicated (aliased) thereafter so that its complex spectral response $H(m)$ is discretised to the microtone resolution for $|m| \leq 0.5(N_\epsilon-1)$. The adjacent microtones won't be orthogonal since the cycle time T is smaller than the microtone period $T_\epsilon$ by a factor $\sigma^{-1}$. This causes spectral leakage from microtones adjacent to $f_b$. \\

Treating equation~(\ref{eq:E_b2}) as if it were an $N_\epsilon$ point DFT spectrum at the microtone resolution, centered at the microtone $b\sigma$ and substituting equation (\ref{eq:E_x_t_2}) in it, the spectral samples at the DFT resolution $f_{res}$ appear as a discrete convolution between the microtone field spectral samples and the window's spectral response interpolated to the microtone resolution:   \\

\begin{eqnarray}\label{eq:dft_2}
E(x, b) &=& \sum_{m=-0.5(N_{\epsilon}-1)}^{0.5(N_{\epsilon}-1)} \sigma E_{m}(x).H(b\sigma-m) \nonumber \\ 
\end{eqnarray}
\\

Let $b+$ and $b-$ indicate positive and negative valued bin indices. The DFT process ensures conjugate symmetry with respect to $b$ since the sign of the microtones follows that of $b$:

\begin{eqnarray}
E(x, b+) &=& \sum_{m=(K+0.5)\sigma+1}^{(K+B_{max}+0.5)\sigma} \sigma E_{m}(x).H(b\sigma-m) \\ \nonumber \\ \nonumber \\
E(x, b-) &=& \sum_{m=-(K+0.5)\sigma+1}^{-(K+B_{max}+0.5)\sigma} \sigma E_{m}(x).H(b\sigma-m)  \nonumber \\
&=& \sum_{m^{\prime}=(K+0.5)\sigma+1}^{(K+B_{max}+0.5)\sigma} \sigma E_{-m^{\prime}}(x).H(-(|b|\sigma-m^{\prime}))  \nonumber \\ \nonumber \\
&=& E^{*}(x, b+) \nonumber \\
\end{eqnarray}
\\

\subsection{Fixed baseline visibility for a point-source \label{subsec:fixed baseline}}
Consider the case where there is no inter-channel migration i.e., having access to visibility at the same baseline $u_{0}$ for all bin-center frequencies ($f_b$) either from  an errorless gridding interpolation or from dense enough packing of antenna elements. In such a scenario, the chromaticity of baselines is from within the spectral channel. \\

The visibility at a baseline $u_0$ for the $b^{th}$ bin is measured by correlating (over x with a shift of $u_0\lambda_b = u_{0}C/(|b|f_{res})$) the radiation field's DFT spectrum for that bin and then time-averaging this over several DFT cycles.

\begin{eqnarray}\label{eq:V_u0}
V_{u_0}(b+) &=& V_{u_0}^{*}(b-) \nonumber \\
&=& \left\langle \; \left\langle E(x, b+).E^{*}((x+u_0\lambda_b), b+)\right\rangle_{x}  \; \right\rangle_{t}  \nonumber \\ \\ \nonumber \\
&=& \sum_{(K+0.5)\sigma+1}^{0.5 N_{\epsilon}} \sum_{(K+0.5)\sigma+1}^{0.5 N_{\epsilon}} \{  \left\langle \; \left\langle E_{m_1}(x).E^{*}_{m_2}(x+u_0\lambda_b) \right\rangle_{x} \;  \right\rangle_{t} \nonumber \\
&.& \sigma^2 H({b\sigma-m_1}) H^{*}({b\sigma-m_2}) \; \} \nonumber \\
\label{eq:V_u0_1}
\end{eqnarray}
\\

Let $L(x)$ be the aperture illumination or weighting function and its spatial auto-correlation, the spatial transfer function, be $W_{u_0}$. For a point-source at $l=l_0$,  \\

\begin{eqnarray}
E_{m}(x) &=&  L(x).E^{src}_m.\exp\left(\frac{\jmath 2\pi l_0 x m \epsilon}{C}\right) 
\end{eqnarray}
\\

$E^{src}_m$, the field emitted by the source at a microtone frequency $m\epsilon$, is a random phase process whose mean amplitude-squared gives the source's power spectral density (per unit microtone spacing) at this microtone frequency namely $P_m$ (equation~\ref{eq:p_analog}). Since it is considered to be of zero-mean and statistically independent for distinct microtones, $\left\langle E^{src}_{m_1}.(E^{src}_{m_2})^{*} \right\rangle_{t} = P_{m_1} \delta_{m_1, m_2}$.

\begin{eqnarray} 
\left\langle \; \left\langle E_{m_1}(x).E^{*}_{m_2}(x+u_0 \lambda_b) \right\rangle_{x} \;  \right\rangle_{t} \nonumber \\
 = P_{m_2} W_{u_0} \exp\left(\frac{-\jmath 2 \pi l_0 m_2 u_0}{|b| \sigma}  \right) \delta_{m_1, m_2} \nonumber \\
\end{eqnarray}
\\

The term $m u_0/(|b| \sigma)$ can be interpreted as the modified baseline due to intra-channel migration at the micro-tone level. At the channel center frequency $f_b$ to which the measured visibility is referenced, let $x_b$ be the antenna separation that would provide the baseline $u_0 = x_b/\lambda_{b} = x_b |b|\sigma\epsilon/C$. The same antenna separation provides a modified baseline $u_m$ at any other microtone frequency ($m\epsilon$):

\begin{equation} \label{eq:um}
u_m = \; \frac{x_b m \epsilon}{C} \; =  \; u_0 \left( \frac{m}{|b|\sigma} \right)
\end{equation}
\\  

Then we can compare the measured visibility from the modified baseline, $V_{u_m}(b)$ against the visibility for the ideal case, $V_{u_0}(b)$ where the baseline would have been maintained at $u_0$ at the microtone level. \\

\begin{eqnarray} 
V_{u_m}(b+) &=& W_{u_0}. \sum_{m_1=(K+0.5)\sigma+1}^{(K+B_{max}+0.5)\sigma} \{ P_{m_1}  \nonumber  \\
&.& \sigma^2 \; |H(|b\sigma-m_1|)|^2 \; \exp\left(\frac{-\jmath 2 \pi l_0 m_1 u_0}{|b| \sigma}  \right) \}
\nonumber \\ \nonumber \\
&=& W_{u_0}. \exp\left(-\jmath 2 \pi l_0 u_0 \right) \sum_{m_1=(K+0.5)\sigma+1}^{(K+B_{max}+0.5)\sigma} \{ \sigma^2 P_{m_1}  \nonumber  \\
&.& |H(|b\sigma-m_1|)|^2 \; \exp\left(\frac{\jmath 2 \pi (b\sigma - m_1) l_0 u_0}{|b| \sigma}  \right) \} \nonumber \\ \nonumber \\
&=& W_{u_0}.\exp\left(-\jmath 2 \pi l_0 u_0 \right).P_{u_m}(m=b\sigma) \nonumber \\
 \label{eq:V_um}  
\end{eqnarray}
\\

\begin{eqnarray}
V_{u_0}(b+) &=& W(u_0).\exp\left(-\jmath 2 \pi l_0 u_0 \right) \nonumber \\
&.&  \sum_{m_1=(K+0.5)\sigma+1}^{(K+B_{max}+0.5)\sigma} \{ P_{m_1} \sigma^2 \; |H(|b\sigma-m_1|)|^2 \; \}  \nonumber \\ \nonumber \\
&=& W(u_0).\exp\left(-\jmath 2 \pi l_0 u_0 \right).P_{u_0}(m=b\sigma)
\nonumber \\ \label{eq:V_u0_2}  
\end{eqnarray}
\\
 
$P_{u_m}(m)$ and $P_{u_0}(m)$ can be interpreted as the power spectral response at a channel centered around $m$, with and without intra-channel migration respectively. $H(m)$ can be taken to have a significant spectral response only till a few (say, $B_{sig}$) channel widths i.e., $H(m) = 0$ for $|m| \geq B_{sig}\sigma$. For reasons like aliasing, the useful signal band $f_{band} =  (f_{min}, \: f_{max}]$ is made a subset of the system band $f_{sys} = (f_{start}, \: f_{end}]$. If $f_{band}$ can be confined to bins with $|b| > (K+B_{sig})$ and $|b| \leq  ((K+B_{max}-B_{sig}) \; = (K+B_{max-sig}))$, then (using $\star$ for discrete convolution),  \\

\begin{eqnarray} 
P_{u_m}(m) &=& \sigma P_{m} \star \{ \sigma |H(m)|^2  \; \exp\left(\frac{\jmath 2 \pi m \epsilon l_0 u_0}{|b| f_{res}} \right) \} \label{eq:P_um} \nonumber \\ \\ \nonumber \\ 
P_{u_0}(m) &=& \sigma P_{m} \star \sigma |H(m)|^2 \label{eq:P_u0}  
\end{eqnarray}
\\

\begin{figure}[!t]
\includegraphics[width=\columnwidth]{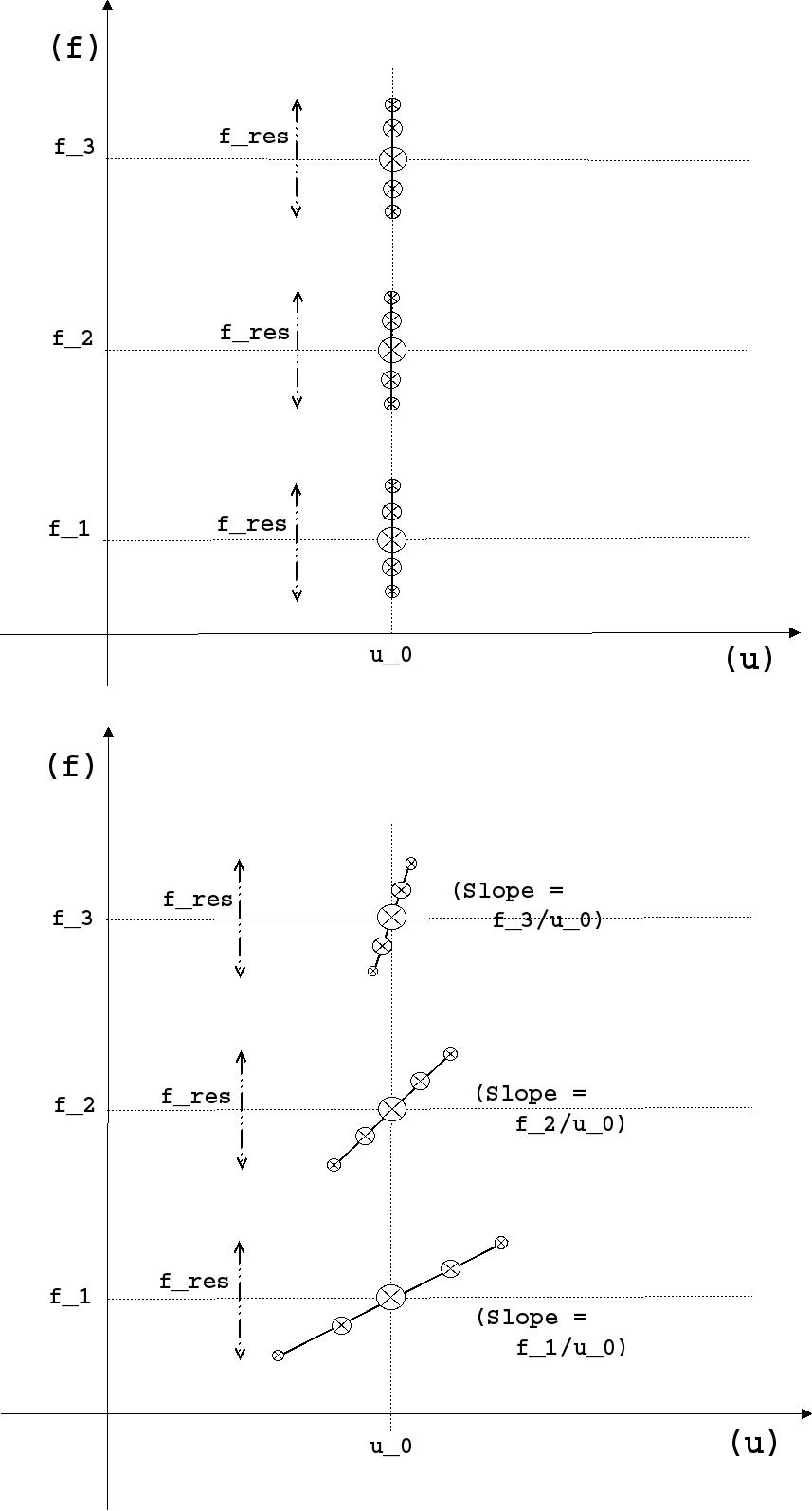}
    \caption{The top and bottom figures show the post-gridding $(u, f)$-plane samples for three arbitrary spectral channels $(f_1 < f_2 < f_3)$ at the reference baseline $u_0$ for the ideal and intra-channel migrated visibilities respectively. The visibilities along the solid line get superposed with a weighting given by the channel's power spectral response, $|H(f-f_b)|^2$ whose strength is depicted by the size of the crossed-circles. In the bottom plot, the solid lines appear to converge at the origin since they lie on position vectors (from origin) to their reference points, $(u_0, f_b)$. } \label{fig:visib_plane}
\end{figure}

In the (u, f)-plane where the visibility measurements are made (Morales M. F. {\em et al.} 2012), this can be interpreted as in figure \ref{fig:visib_plane}. In the ideal case, the same phase of the visibility gets sampled for all microtones that go into a channel. In the measured case, there is a shift in the phase of the visibility sampled at one microtone compared to its neighbour and this shift increases as the ratio of the channel center frequency to the spectral resolution (i.e., channel index) decreases. In other words, the extent (baseline span) of intra-channel migration increases as this ratio decreases. \\

$R_{h}(\tau)$, the inverse transform of $|H(m)|^2$, is the auto-correlation of $h(t)$ normalised to microtone period. We'll use $R^{T}_{h}(\tau)$ to denote the inverse transform of $\sigma |H(m)|^2$ which normalises it to the duration of the window's existence, $T$. 

\begin{eqnarray}
R_{h}(\tau) &=& T_{\epsilon}^{-1} \int_{0}^{T_{\epsilon}} {h(t).h(t-\tau)} dt \\
R^{T}_{h}(\tau) &=& \sigma R_{h}(\tau) = T^{-1} \int_{0}^{T} {h(t).h(t-\tau)} dt
\end{eqnarray}
\\

Using $R_{src}(\tau)$ (equation~\ref{eq:r_analog}) for the source's auto-correlation, the fourier transforms of $P_{u_m}(m)$ and $P_{u_0}(m)$ can be seen as instrumental windowing on the source's auto-correlation:

\begin{eqnarray} 
R_{u_m}(\tau) &=& \sigma R_{src}(\tau).\; R^{T}_{h}\left(\tau+\frac{l_0 u_0 T}{|b|} \right)  \label{eq:R_um} \nonumber \\ \\ \nonumber \\ 
R_{u_0}(\tau) &=& \sigma R_{src}(\tau).\; R^{T}_{h}(\tau) \label{eq:R_u0}  
\end{eqnarray}
\\
 
When there is intra-channel migration, the auto-correlation peaks of the source and the window function are not aligned anymore. This attenuates the measured visibility. The mis-alignment $(l_0 u_0 T)/(|b|)$ is frequency dependent and keeps increasing as $b$, the ratio of channel center frequency to the resolution decreases in magnitude. \\

\subsection{Quantifying the intra-channel artifact for flat-spectral density \label{subsec:flat-spectrum}}

Here $P_m = P_{l_0}$ is constant and $R_{src}(\tau) = P_{l_0} \delta (\tau)$ (equations \ref{eq:p_analog} and \ref{eq:r_analog}). Let $Q_{l_0} \; = \; \sigma P_{l_0}$ be the  power spectral density from the pixel $l_0$ per unit channel spacing. Let the normalization to $h(t)$ be such that, $R^{T}_{h}(0)$, its power content within $0 \leq t \leq T$ is unity. This makes 

\begin{eqnarray} 
P_{u_0}(m) &=& Q_{l_0} \label{eq:P_u0_2} \\
P_{u_m}(m) &=& Q_{l_0} R^{T}_{h}\left(\frac{|l_0 u_0| T}{|b|} \right)  \label{eq:P_um_2} 
\end{eqnarray} 
\\

Due to equation~(\ref{eq:V_um}), the manner in which the measured visibility decays spectrally at a channel $b$ depends on how $R^{T}_{h}(\tau)$ decorrelates at a lag of $(|l_0 u_0|T/|b|)$. Since the latter exists only for $|\tau| < T$ and $(nT_{\epsilon}-T) < |\tau| <  (nT_{\epsilon}+T)$, where $n = 1, 2,... \infty $ denotes the aliasing index, the former exists only if the channel index, $|b| = (|f_{b}|/f_{res})$ lies within a certain range $(R^0 \cup R^n)$. \\

The region $R^0$ is $(|l_0 u_0|, \; \infty)$ and $R^n$ is $$\left(\frac{|l_0 u_0|}{n\sigma+1}, \; \frac{|l_0 u_0|}{n\sigma-1}  \right)$$ \\

Let $\Delta R^0$ and $\Delta R^n$ denote the range of bins provided by the corresponding regions over which the visibility does not vanish. Given that $\sigma >> 1$, 
 
\begin{eqnarray} \label{eq:Delta_Rn}
\Delta R^n &=& \sum_{n=1}^{\infty} \left\lbrace \frac{|l_0 u_0|}{n\sigma-1} - \frac{|l_0 u_0|}{n\sigma+1}  \right\rbrace \\
&\approx& 2 \frac{|l_0 u_0|}{\sigma^2} \sum_{n=1}^{\infty} \frac{1}{n^2} =  \frac{|l_0 u_0|}{\sigma^2} \frac{\pi^2}{3} 
\end{eqnarray}
\\

The per-baseline worst-case (maximum) for $|l_0 u_0|$ is 

\begin{eqnarray} 
M_{u_0} &=& l_{max}|u_0| \; = \; 0.5 \; |u_0|.\frac{\lambda_{min}}{x_{min}} \label{eq:M_u0} \\  \nonumber \\
M_{u_{max}} &=& l_{max}u_{max} \; = \; 0.5 \; \frac{x_{max}}{x_{min}}.\frac{\lambda_{min}}{\lambda_{max}} \nonumber \\
&=& \; 0.5 \; \frac{x_{max}}{x_{min}}.\frac{f_{min}}{f_{max}} \nonumber \\ \label{eq:M_umax}
\end{eqnarray}
\\

Those ranges of $R^n$ that fall below $(K+B_{sig})$ do not count. Moreover $\Delta R^n$ is less than one and the frequency range provided by $R^n$ over all n does not exceed $f_{res}$ so long as $(\pi^2/3).(M_{u_{max}}/\sigma^2)<1$ and so at most one spectral channel can have $|b|$ within $R^n$. We ignore this and consider only $R^0$ as the valid range for $|b|$. \\

 $R^0$ can be rewritten as $((K+B_{crit}) < |b| \leq (K+B_{max-sig}))$ where,

\begin{eqnarray} \label{eq:B_crit_def}
B_{crit}(u_0) &=& Max\left\lbrace B_{sig}, \; (|l_0 u_0| - K)\right\rbrace 
\end{eqnarray}
\\

 For this to hold in the entire field-of-view, $-l_{max} \leq l_0 \leq l_{max}$, we need:

\begin{eqnarray} \label{eq:B_crit}
B_{crit}(u_0) &=& Max\left\lbrace B_{sig}, \; (M_{u_0} - K)\right\rbrace 
\end{eqnarray}

\begin{eqnarray} \label{eq:Delta_R0}
\Delta R^0 &=&  (B_{max-sig} - B_{crit}(u_0))
\end{eqnarray}
\\

We'll proceed by posing the visibility attenuation as a function of the channel center  frequency. 

\begin{eqnarray}
V_{u_m}(b) &=& G_{u_0}(b).V_{u_0}(b) \label{eq:V_um_5}
\end{eqnarray}
\\

\begin{eqnarray}
G_{u_0}(b) &=& R^{T}_{h}\left(\frac{|l_0 u_0| \; T}{|b|} \right), \nonumber \\
&for& \; (B_{crit} < (|b|-K) \leq B_{max-sig}) \nonumber \\ 
&or& \; 0, \; otherwise \nonumber \\
\label{eq:G}
\end{eqnarray}
\\

\begin{figure}[!t]
\includegraphics[width=\columnwidth]{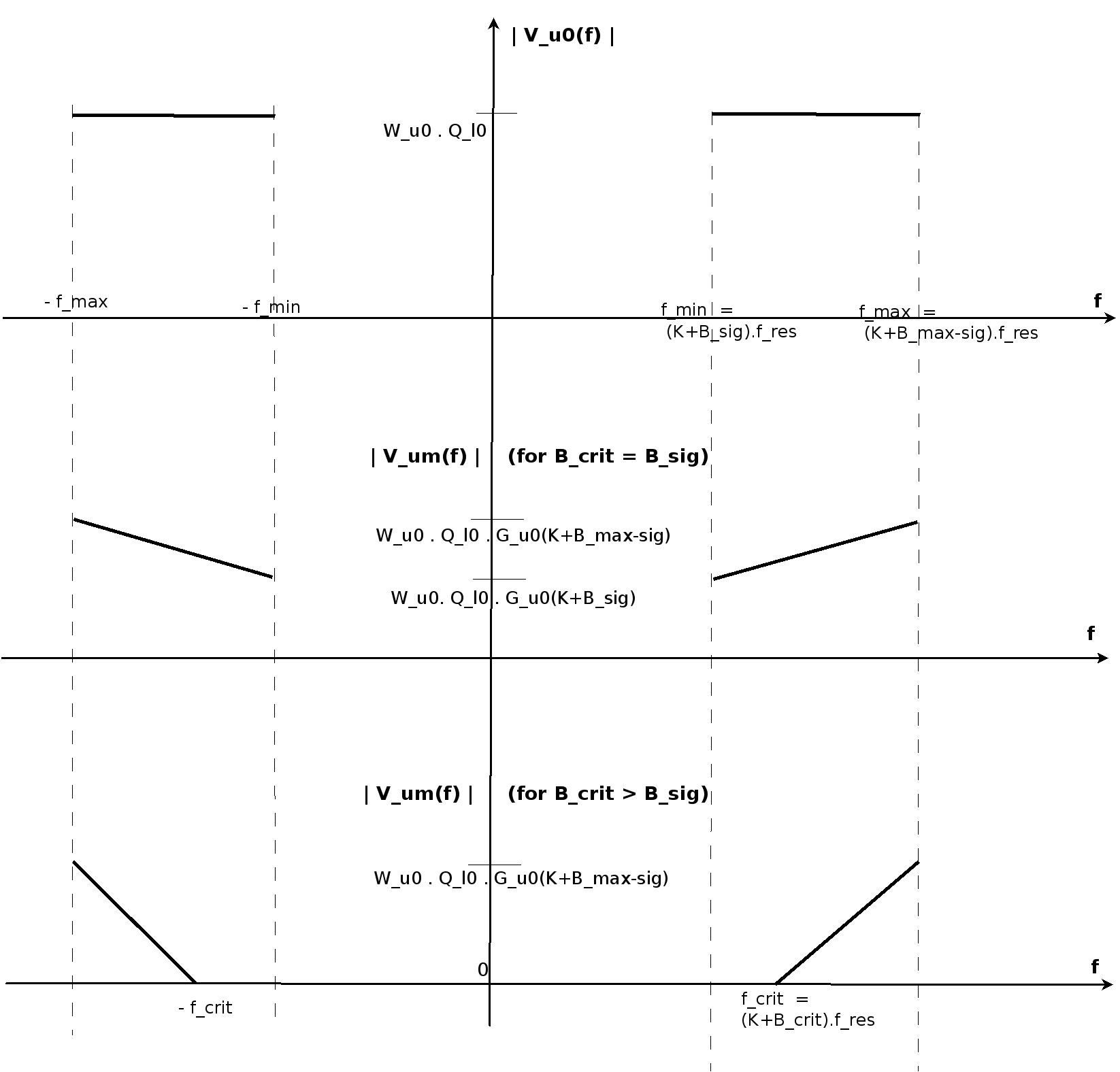}
    \caption{Spectral dependence of visibility magnitude for a flat spectrum point source. The top plot shows the magnitude of the ideal visibility $V_{u_0}(f=bf_{res})$ across the band. The bottom two plots show the measured case, $V_{u_m}(f=bf_{res})$, for $B_{crit} = B_{sig}$ and for $B_{crit} > B_{sig}$ respectively.} \label{fig:visib_freq}
\end{figure}

The sub-unity factor $G$ is maximum for the highest frequency and falls gradually to a  minimum at $|b| = (K+B_{crit})$. This minimum is zero if $B_{crit} > B_{sig}$ where the former equals ($|l_0 u_0| - K$). Figure~\ref{fig:visib_freq} shows the magnitude of visibility as a function of frequency. \\

For the natural DFT window, $h(t) = 1 \; for \; 0 \leq t \leq T$,  

\begin{eqnarray} \label{eq:A_natural}
R^{T}_{h}(\tau) &=& 1 - \frac{\tau}{T}, \; if \; |\tau| \leq T \nonumber \\
&or& 0, \; if \; T \leq |\tau| \leq T_{\epsilon}  
\end{eqnarray}
\\

\begin{eqnarray}
G_{u_0}(b) &=& \left(1-\frac{|l_0 u_0|}{|b|} \right), \nonumber \\
&for& \; (B_{crit}(u_0) < (|b|-K) \leq B_{max-sig})  \nonumber \\ 
&or& \; 0, \; otherwise \nonumber \\
\label{eq:G_natural}
\end{eqnarray}
\\

The average slope of decay across the band is,

\begin{eqnarray}
\Delta_{G} &=& \frac{G_{u_0}(K+B_{max-sig})-G_{u_0}(K+B_{crit})}{B_{max-sig} - B_{crit}}
\nonumber \\ \nonumber \\
&=& \frac{|l_0 u_0|}{(K+B_{max-sig}).(K+B_{crit})}  \\ \nonumber \\
&=& |l_0 u_0| \; \frac{f^{2}_{res}}{f_{max} f_{crit}}  
\label{eq:G_slope}
\end{eqnarray}
\\

If $(|b|-K) < B_{max-sig} \ll K$ i.e., $f_{band} \ll f_{start}$, we have this linearization, $G_{u_0}(b) \approx G^{base} + G^{sloping}(b)$ such that for $B_{crit} < (|b|-K) \leq B_{max-sig}$,

\begin{eqnarray} \label{eq:G_base}
G^{base} &=& G_{u_0}(|b|=(K+B_{crit}+1)) \nonumber \\
&=& G_{u_0}(K+B_{sig}+1), \; if \; B_{crit} = B_{sig}  \nonumber \\
&or& \Delta_G, \; if \; B_{crit} > B_{sig} \nonumber \\
\end{eqnarray}

\begin{eqnarray} \label{eq:G_sloping}
G^{sloping}(b) &=&  \Delta_{G}.(|b| - (K+B_{crit}+1))  \nonumber \\
\end{eqnarray}
\\

For $B_{crit} \ngtr B_{sig}$, the amplitude of the sloping term relative to the base term is,

\begin{eqnarray} \label{eq:G_ratio}
\frac{G^{sloping}(K+B_{max-sig})}{G^{base}}  &=&  \frac{|l_0 u_0|}{\left(1 + \frac{f_{min}}{f_{res}} - |l_0 u_0| \right)} . \frac{f_{band}}{f_{max}}  \nonumber  \\
\end{eqnarray}
\\

As the spectral resolution worsens, $\sigma$ increases while $K$ decreases and the slope of G increases. When $(K+B_{sig}) = f_{min}/f_{res}$ drops below $|l_0 u_0|$, $B_{crit}(u_0)$ exceeds $B_{sig}$ and  $V_{u_m}$ starts to shrink from its original span of $f_{band}$ and in the delay domain this stretches the delay spectrum as shown in section~\ref{sec:Section3}. This crossing first happens at the highest baseline and thereafter the extent of shrinking $(B_{crit}(u_0) - B_{sig})$ increases with baseline.  \\

\subsection{Using polyphase filter bank \label{subsec:PFB}}
A polyphase filter bank (PFB) (Chennamangalam J. 2016; Price D. C. 2018; Thompson, A. R., Moran, J. M., Swenson, Jr., G. W. 2017) can be seen as a window that operates over several cycles (say, $P$) of buffered data, thereby shrinking the inter-channel spectral leakage by P (i.e., $H(m)$ becomes $H(mP)$ and the channel resolution improves from $f_{res}$ to $f_{res}/P$). Now there are $P$ times more channels to cover the same signal band and we decimate the DFT output taking every $P^{th}$ channel so that the spacing between adjacent  channels is still kept at  $f_{res}$. Then equations~(\ref{eq:E_b}) and (\ref{eq:E_b2}) become,

\begin{eqnarray}
E(x, Pb) &=& \frac{1}{PT/T_s} \sum_{n=0}^{PT/T_s} \{ h(t=nT_s).E(x, t=nT_s) \nonumber \\ &.&\exp( -\jmath 2 \pi f_{b} nT_s) \}  
\nonumber \\ \\ \nonumber \\ \label{eq:E_b_pfb} 
&=& \frac{\sigma}{P N_\epsilon} \sum_{n=0}^{(N_\epsilon-1)} h(t=nT_s).E(x, t=nT_s)
\nonumber \\ &.&\exp( -\jmath 2 \pi f_{b} nT_s)   
\nonumber \\ \label{eq:E_b2_pfb}  
\end{eqnarray}
\\

where the first $((P-1)T/T_s)$ time samples come from buffered data of the previous $(P-1)$ DFT cycles and the last $T/T_s$ samples are the fresh data input for the current cycle. $P$ is the order (number of taps) of polyphase FIR filters used in the filter bank. Since the PFB window function is $P$ times larger compared to a window operating on an isolated cycle of DFT input (also referred to as a single block fourier transform (SBFT)), $R^{T}_{h}(\tau)$ remains non-zero for $|\tau| < PT$ as opposed to $T$. This relaxes the region $R^0$ to $(|l_0 u_0|/P, \; \infty)$. \\

 Although spectral leakage from microtones outside the channel has been reduced relative   to SBFT, the microtones from the same channel cause the measured visibility to decay and possibly vanish if $M_{u_{max}}/(K+B_{sig}) = M_{u_{max}}.f_{res}/f_{min}$ exceeds $P$.  \\ \\

\subsection{Simulating for a distribution of flat-spectrum point sources \label{subsec:sim_pt_src_dist}}
\begin{figure*}[!t]
\centering
\includegraphics[width= 2.5\columnwidth]{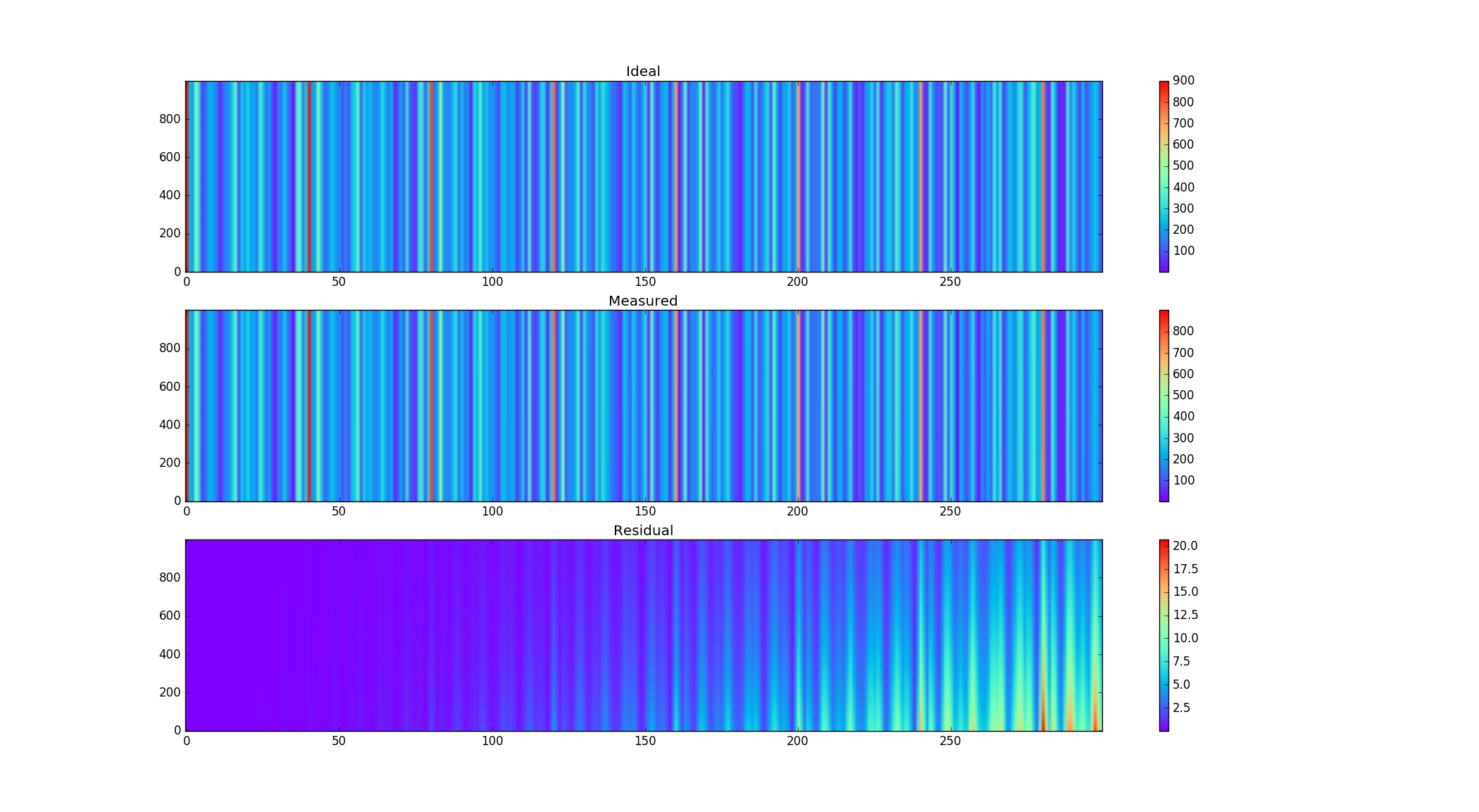}
    \caption{For a distribution of flat-spectrum point sources, the visibility magnitude in the $u-f$ plane is shown for the ideal, intra-channel migrated and the residual cases respectively. The horizontal and vertical axes are marked respectively with the baseline and channel indices. The colormap used is $rainbow$. It is clear in the residual plot that the intra-channel artifact becomes more severe as we go from left to right (baseline increases) or from top to bottom (channel center frequency decreases). } \label{fig:visib_sim}
\end{figure*}

Figure \ref{fig:visib_sim} shows simulated results from a distribution of flat-spectrum point sources. The following system parameters were used: signal band in $300$ to $500 \; MHz$, spectral resolution of $200 \; kHz$, minimum and maximum antenna separations of $x_{min} = 1.5$ and $x_{max} = 748.5$ meters respectively. The minimum and maximum reference baselines common to all channels are taken to be $u_{min} = x_{min}/\lambda_{min} = 2.502$ and $u_{max} = x_{max}/\lambda_{max} = 749.018$ respectively. The intermediate baselines are integer multiples of $u_{min}$. \\ 

For the channel's spectral response, a single-block Blackman-Nuttal window with coefficients as specified in Table-2 of Price D. C. (2018), was used.  The microtone resolution was set to $200 \; Hz$ so that $\sigma = 1000$.  To simulate the intra-channel effect, visibilities from microtones spanning $\pm 5 f_{res}$ around the channel center frequency were weighted by $|H(f-f_b)|^2$ and superposed. For a single point source, the magnitude of residual visibility $|1-G_{u_0}(b)|.|V_{u_0}(b)|$ is a function of $|l_0 u_0|f_{res}/|f_b|$. For the point source distribution, as seen in the third plot, the residual  becomes worse as $|u_0|$ increases or $|f_b|$ decreases. \\ \\

\section{Effect on the spatial power spectrum} \label{sec:Section3}
$V_u(\eta)$ is obtained by transforming (from $f$ to $\eta$) the visibilities at a constant reference baseline $u$. The spatial power spectrum is given by $|V_u(\eta)|^2$ with $u$ and $\eta$ being proportional to the transverse $(k_{\perp})$ and line-of-sight $(k_{\parallel})$ wave numbers respectively. This is to be contrasted with the delay spectrum, $V_x(\tau)$, (Parsons A. R. {\em et al.} 2012) which transforms (from $f$ to $\tau$) the visibilities at a constant antenna separation of $x$ meters. See Thyagarajan N. {\em et al.} (2015a) for a comparison between $V_u(\eta)$ and $V_x(\tau)$. \\
 
$V_x(\tau)$ is a simplification and approximation to $V_u(\eta)$ which gives the actual fourier transform of the cosmological data cube. Since the former uses the measured visibility samples as such without any gridding, inter-channel migration is inherent in it besides intra-channel migration with a constant $u-f$ plane slope of $C/x$ at every channel.  Here, as indicated in the beginning of section \ref{subsec:fixed baseline}, we consider what happens when inter-channel migration has been eliminated and we use only $V_u(\eta)$. \\

\begin{eqnarray}  
V_u(\eta) &=& \int_{-\infty}^{\infty} V_u(f) \exp(-\jmath 2 \pi \eta f) df
 \label{eq:V_u_eta}
\end{eqnarray}
\\

$V_u(f)$ is sampled by our measurement for $f = bf_{res}$ with $|b| \leq (K+B_{max-sig})$  and replicated (aliased) outside this range with the spectral period $2f_{max}$.

\begin{eqnarray}
N_{sig} &=& \frac{2 f_{max}}{f_{res}} = 2(K+B_{max-sig}+0.5)
\end{eqnarray}
\\

Consequently, the $\eta$ domain is sampled at a delay spacing of $\eta_s = (2f_{max})^{-1}$ and replicated with the delay period $(f_{res})^{-1}$. Let $\eta = d \eta_s = d / (N_{sig} f_{res})$ denote the delay samples for the integer $d$ in a continuous range of $N_{sig}$ samples. \\

\begin{eqnarray}  
V_u(d) &=& \sum_{b=-(K+B_{max-sig})}^{(K+B_{max-sig})} V_u(b) \exp\left(\frac{-\jmath 2 \pi b d}{N_{sig}}\right) \nonumber \\
 \label{eq:V_u_eta_2}
\end{eqnarray}
\\

The ideal visibility from equations (\ref{eq:V_u0_2}) and (\ref{eq:P_u0_2}) is,

\begin{eqnarray}  
V_{u_0}(b) &=& Q_{l_0} W_{u_0}. \exp(-\jmath 2 \pi l_0 u_0), \nonumber \\
&for& \; ((K+B_{sig}) < b \leq (K+B_{max-sig})) \nonumber \\ \nonumber \\
&or& Q_{l_0} W_{u_0}. \exp(\jmath 2 \pi l_0 u_0), \nonumber \\
&for& \; (-(K+B_{max-sig}) \leq b < -(K+B_{sig})) \nonumber \\ \nonumber \\
&or& \; 0, \; elsewhere \nonumber \\
 \label{eq:V_u0_5}
\end{eqnarray}
\\

\begin{eqnarray}  
V_{u_0}(d) &=& Q_{l_0} W_{u_0}. 2 B_{diff} \frac{sinc(B_{diff} d/N_{sig})}{sinc(d/N_{sig})} \nonumber \\
&.& \cos\left(\frac{2 \pi d (K+B_{av})}{N_{sig}} + 2 \pi l_0 u_0 \right)  \nonumber \\
 \label{eq:V_u0_d}
\end{eqnarray}
\\ 

\noindent where $B_{diff} = (B_{max-sig} - B_{sig})$ and $B_{av} = (B_{max-sig} + B_{sig} + 1)/2$. The ratio of sincs (Dirichlet kernel) signifies replication in $\eta$-domain due to sampling in $f$-domain. The numerator sinc (of a delay spread of $(N_{sig}/B_{diff})$ delay samples) is replicated at a period (of $N_{sig}$ samples) given by the denominator sinc. \\ 

For the natural window (equation~\ref{eq:G_natural}), the base and sloping terms (equations \ref{eq:G_base} and \ref{eq:G_sloping}) when multiplied with $V_{u_0}(b)$ and delay transformed give distinct delay modes. The base term being flat across the band, gives a delay mode similar to the ideal one in equation~(\ref{eq:V_u0_d}). With $B^{\prime}_{diff} = (B_{max-sig} - B_{crit})$ and $B^{\prime}_{av} = (B_{max-sig} + B_{crit} + 1)/2$,

\begin{eqnarray}  
V_{u_m}^{base}(d) &=& Q_{l_0} W_{u_0}.G^{base}. 2 B^{\prime}_{diff} \frac{sinc(B^{\prime}_{diff} d/N_{sig})}{sinc(d/N_{sig})} \nonumber \\
&.& \cos\left(\frac{2 \pi d (K+B^{\prime}_{av})}{N_{sig}} + 2 \pi l_0 u_0 \right)  \nonumber \\
 \label{eq:V_um_base_d}
\end{eqnarray}
\\

\begin{eqnarray}  
&V&_{u_m}^{sloping}(d) \nonumber \\
&=& 2 B^{\prime}_{diff}Q_{l_0}W_{u_0}.\Delta_G.\frac{B^{\prime}_{diff}-1}{2}.\cos\left(2 \pi l_0 u_0 \right), \nonumber \\ 
&if& \; d = 0 \; or \; integer \; multiple \; of \; N_{sig}  \nonumber \\ \nonumber \\
&or& B^{\prime}_{diff}Q_{l_0}W_{u_0}.\Delta_G.\{ (-1)^d \frac{sin(2 \pi l_0 u_0)}{sin(\pi d / N_{sig})} \nonumber \\
&-& \frac{sin(2 \pi l_0 u_0)}{sin\left(\frac{\pi d}{N_{sig}}\right)} \cos\left(\frac{2 \pi d (K+B^{\prime}_{av})}{N_{sig}}\right) \frac{sinc\left(\frac{B^{\prime}_{diff} d}{N_{sig}}\right)}{sinc\left(\frac{d}{N_{sig}}\right)} \nonumber \\
&-& 2 (K + B^{\prime}_{av}) \cos(2 \pi l_0 u_0) \frac{sinc\left(\frac{B^{\prime}_{diff} d}{N_{sig}}\right)}{sinc\left(\frac{d}{N_{sig}}\right)} \nonumber \\
&.& \frac{sinc\left(\frac{2 (K+B^{\prime}_{av}) d}{N_{sig}}\right)}{sinc\left(\frac{d}{N_{sig}}\right)} \}, \; elsewhere \nonumber \\
 \label{eq:V_um_sloping_d}
\end{eqnarray}
\\

There are three terms being summed. In terms of amplitude they all are weak, of the order of the base delay mode when $B_{crit} > B_{sig}$. In terms of delay spread, the last two terms aren't any worse than the base mode but the first persists over the entire delay range. It decays as $cosec(\pi d / N_{sig})$ which remains greater than unity in $|d| < 0.5 N_{sig}$. This persistent delay mode with an amplitude $\sim 0.5 \Delta_G B^{\prime}_{diff}/B_{diff}$ times that of the ideal delay mode is solely an effect of intra-channel migration. So is the attenuation of the base mode by the factor $G^{base}$. \\

\subsection{Reduction in bandwidth and stretching of delay mode when $B_{crit}$ exceeds $B_{sig}$  \label{subsec:Delay_stretch}}

Since the delay spread is inversely related to the pass band width (barring the persistent mode), a per-baseline stretching factor can be defined from its ratio for the ideal to the measured case:

\begin{eqnarray} \label{eq:S}
S_{u_0} &=& \frac{B_{diff}}{B^{\prime}_{diff}} = \left( \frac{B_{max-sig} - B_{sig}}{B_{max-sig} - B_{crit}(u_0)}\right) \\ \nonumber \\
&=& \left( \frac{1}{1 - \frac{f_{res}(B_{crit}(u_0) - B_{sig})}{f_{band}}}\right) \\ \nonumber \\ \nonumber \\
&=& 1, \; if \; \; (B_{crit} = B_{sig}) \nonumber \\
&or& \left(\frac{1}{1 - \frac{0.5|u_0|f_{res}\lambda_{min}/x_{min} - f_{min}}{f_{band}}} \right), \nonumber \\ \nonumber \\ 
&if& \; \; (B_{crit} > B_{sig}) \: i.e., \: (M_{u_0} > f_{min}/f_{res}) \nonumber \\ \\ \nonumber \\
&=& \; \; 1 + \left(\frac{0.5|u_0|f_{res}\lambda_{min}/x_{min} - f_{min}}{f_{band}} \right),
\nonumber \\ \nonumber \\
&for& \; \; (B_{sig} < B_{crit} \ll B_{max-sig}) \nonumber \\
\end{eqnarray}
\\

We see that for $B_{crit} > B_{sig}$, the extent of stretching increases with the baseline. Viewed in the instrument k-space, the lower-limit of $|k_{\parallel}|$ is raised (from the expected value of the order of reciprocal bandwidth) by the factor $S_{u_0}$ which increases with $|k_{\perp}| \sim |u_0|$. When $B_{crit}$ touches $(B_{max-sig} - 2)$ for a particular $|k_{\perp}|$ mode, the lower limit of $|k_{\parallel}|$ coincides with the upper limit defined by $0.5(f_{res}^{-1})$ making the k-space unusable from that $|k_{\perp}|$ mode onwards. \\ 

This stretching starts when $B_{crit}$ grows from $B_{sig}$ and then becomes infinite (i.e., leaks from zeroth mode to all higher ones) as $B_{crit}$ approaches $B_{max-sig}$. These two conditions provide primary and secondary limits for the DFT spectral resolution.  The primary limit $B_{crit} \ngtr B_{sig}$ applied to $u_0 = u_{max}$ implies $M_{u_{max}} \leq ((K+B_{sig}) = f_{min}/f_{res})$ and

\begin{eqnarray} 
f_{res} &\leq& 2 \;  f_{max}.\frac{x_{min}}{x_{max}}\label{eq:f_res_prim} 
\end{eqnarray}
\\ 

The secondary limit $B_{crit} < B_{max-sig}$ applied to $u_0 = u_{max}$ implies $M_{u_{max}} < ((K+B_{max-sig}) = f_{max}/f_{res}$ and

\begin{eqnarray} 
f_{res} &<& 2 \; \frac{f_{max}^2}{f_{min}}.\frac{x_{min}}{x_{max}}\label{eq:f_res_sec} 
\end{eqnarray}
\\

For a two-dimensional array, $l_0 u_0$ gets replaced by $(l_0 u_0 + m_0 v_0)$ for source position $(l_0, m_0)$ and the reference baseline $(u_0, v_0)$. In the $(u, v, f)$ space, the solid line (similar to the bottom plot in figure \ref{fig:visib_plane}) depicting intra-channel migration around a reference point $(u_0, v_0, f_b)$ lies on the position vector (from origin) to this point. \\

\noindent Equations (\ref{eq:M_u0}) and (\ref{eq:M_umax}) become

\begin{eqnarray} 
M_{u_0, v_0} &=& (l_{max}|u_0| + m_{max}|v_0|) \nonumber \\
&=& 0.5 \; \left(|u_0|.\frac{\lambda_{min}}{x_{min}} + |v_0|.\frac{\lambda_{min}}{y_{min}}\right) \label{eq:M_u0_2d} \\  \nonumber \\
M_{u_{max}, v_{max}} &=& (l_{max} u_{max} + m_{max} v_{max}) \nonumber \\
&=& 0.5 \; \left(\frac{x_{max}}{x_{min}} + \frac{y_{max}}{y_{min}}\right).\frac{\lambda_{min}}{\lambda_{max}} \nonumber \\
&=& \; 0.5 \; \left(\frac{x_{max}}{x_{min}} + \frac{y_{max}}{y_{min}}\right).\frac{f_{min}}{f_{max}}  \nonumber \\ \label{eq:M_umax_2d}
\end{eqnarray}
\\

\balance

The primary and secondary limits become 

\begin{eqnarray} 
f_{res} &\leq& 2 \;  f_{max}.\left(\frac{x_{max}}{x_{min}} + \frac{y_{max}}{y_{min}}\right)^{-1}\label{eq:f_res_prim_2D} 
\end{eqnarray}

\begin{eqnarray} 
f_{res} &\leq& 2 \;  \frac{f_{max}^2}{f_{min}}.\left(\frac{x_{max}}{x_{min}} + \frac{y_{max}}{y_{min}}\right)^{-1}\label{eq:f_res_sec_2D} 
\end{eqnarray}
\\

As indicated in section \ref{subsec:utone}, if IF down-conversion is used in the analog chain, $f_{max}$ and $f_{min}$ should correspond to the original RF band while $f_{res}$ is what is achieved in the IF band. \\ \\

\vspace{-2em}
\section{Conclusion}
We have used the visibility for a flat-spectrum point source to demonstrate the artifacts arising out of a finite spectral resolution, even if the baseline migration between channels has been eliminated. We see that the FX-correlated, per-channel visibility undergoes an overall attenuation. This in turn attenuates the main (base) delay mode.  Moreover there is a gradual decay in visibility from the highest frequency to the lowest. This "sloping" has given rise to a new mode and a part of it persists over the entire delay space. \\

As the spectral resolution worsens, the attenuation as well as the slope (across the band) increase to a point where the visibility can vanish at the lowest frequency. We call this the primary critical limit. At a further secondary limit, it would have vanished for all higher frequencies in the band. These limits relate the channel resolution to the ratio of maximum to minimum antenna separations and the maximum and minimum frequencies for the signal band. A practical interferometer needs to ensure that its spectral resolution is well within these limits to minimise the artifact in visibility caused by intra-channel baseline migration.  \\

\section*{Acknowledgements}
Acknowledge the use of ADS database and the "Dia" diagram editor. \\

\vspace{-1em}



\end{document}